\documentclass[showpacs,preprintnumbers,amsmath,amssymb]{revtex4}
\usepackage{graphicx}
\usepackage{dcolumn}
\usepackage{bm}
\usepackage{amssymb}
\usepackage{array}
\usepackage{hhline}
\usepackage{longtable}

\begin{document}

\title{The rigorous solution for the average distance of a Sierpinski network}

\author{Zhongzhi Zhang$^{1,2}$}
\email{zhangzz@fudan.edu.cn}

\author{Lichao Chen$^{1,2}$}

\author{Lujun Fang$^{3}$}

\author{Shuigeng Zhou$^{1,2}$}
\email{sgzhou@fudan.edu.cn}

\author{Yichao Zhang$^{4}$}

\author{Jihong Guan$^{4}$}
\email{jhguan@tongji.edu.cn}

\affiliation {$^{1}$Department of Computer Science and Engineering,
Fudan University, Shanghai 200433, China}

\affiliation {$^{2}$Shanghai Key Lab of Intelligent Information
Processing, Fudan University, Shanghai
200433, China} %

\affiliation{$^{3}$Electrical Engineering and Computer Science
Department, University of Michigan, 2260 Hayward Ave., Ann Arbor,
Michigan 48109, USA}

\affiliation{$^{4}$Department of Computer Science and Technology,
Tongji University, 4800 Cao'an Road, Shanghai 201804, China}

\begin{abstract}

The closed-form solution for the average distance of a deterministic
network--Sierpinski network--is found. This important quantity is
calculated exactly with the help of recursion relations, which are
based on the self-similar network structure and enable one to derive
the precise formula analytically. The obtained rigorous solution
confirms our previous numerical result, which shows that the average
distance grows logarithmically with the number of network nodes. The
result is at variance with that derived from random networks.

\end{abstract}


\date{\today}
\maketitle

\section{Introduction}

Structural properties~\cite{CoRoTrVi07}, such as degree
distribution~\cite{BaAl99}, average distance~\cite{WaSt98}, degree
correlations~\cite{Newman02}, community~\cite{GiNe02},
motifs~\cite{MiShItKaChAl02}, fractality~\cite{SoHaMa05}, and
symmetry~\cite{XiXiWaWa08}, have received much attention in the
field of complex networks, since these features play significant
roles in characterizing and understanding complex networked systems
in nature and society. Among these important features, average
distance characterizes the small-world behavior commonly observed in
various disparate real networks~\cite{WaSt98}. It has been
established that average distance is related to other structural
properties, such as degree distribution~\cite{ChLu02,CoHa03},
fractality~\cite{SoHaMa06,ZhZhZo07} and
symmetry~\cite{XiMaiWaXiWa08}. On the other hand, average distance
has an important consequence on dynamical processes taking placing
on networks, including disease spreading~\cite{WaSt98},
routing~\cite{YaZhHuFuWa06, ZhCoFeRaRoZh08},
robustness~\cite{SoHaMa06,ZhZhZo07}, percolation~\cite{ZhZhZoCh08},
and so on. Thus far, average distance has become a focus of
attention for the scientific
community~\cite{DoMeSa03,FrFrHo04,Lolo03,HoSiFrFrSu05,DoMeOl06,ZhZhChYiGu08}.

Using above-mentioned structural properties, extensive empirical
studies on diverse real systems have been done with an attempt to
uncover and understand the generic features and complexity of these
systems, and various network models have been proposed to reproduce
or explaining the common characteristics of real-life
networks~\cite{AlBa02,DoMe02,Ne03,BoLaMoChHw06}. Recently, inspired
by the well-known Sierpinski gasket, we proposed a novel network,
called Sierpinski network~\cite{ZhZhFaGuZh07}. The Sierpinski
network belongs to a deterministically growing class of networks
that have attracted considerable attention and have turned out to be
a useful
tool~\cite{BaRaVi01,DoGoMe02,JuKiKa02,RaSoMoOlBa02,AnHeAnSi05,Hi07,BoGoGu08}.
Many relevant topological properties of Sierpinski network such as
degree distribution, clustering coefficient, and strength
distribution have been determined analytically~\cite{ZhZhFaGuZh07}.
Also, the average distance of Sierpinski network has been been
investigated numerically, which was shown to behaves a logarithmic
scaling with the number of network nodes
(vertices)~\cite{ZhZhFaGuZh07}.

In view of the importance and usefulness of the quantity---average
distance, here we derive a closed-form formula for the average
distance characterizing the Sierpinski network. The analytic method
is on the basis of the recursive construction and self-similar
structure of Sierpinski network. Our precise result shows that the
average distance of Sierpinski network increases logarithmically
with the number of nodes. This scaling behaves differently from that
of random networks~\cite{ChLu02,CoHa03}. Our rigorous solution
confirms the scaling between average distance and number of network
nodes that was previously obtained numerically
in~\cite{ZhZhFaGuZh07}.

\begin{figure}
\begin{center}
\includegraphics[width=12cm]{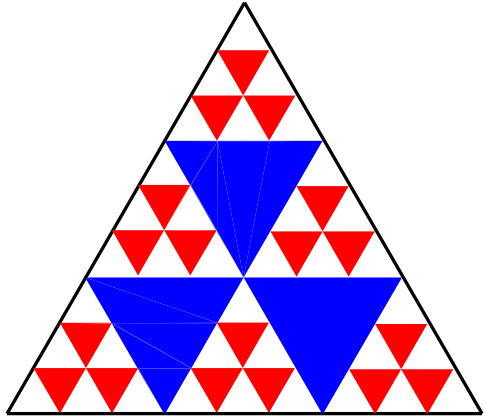}
\end{center}
\caption{\label{Sier} The first two stages in the construction for a
variant of Sierpinkski gasket.}
\end{figure}

\section{Brief introduction to the Sierpinski network}

\begin{figure}
\begin{center}
\includegraphics[width=8cm]{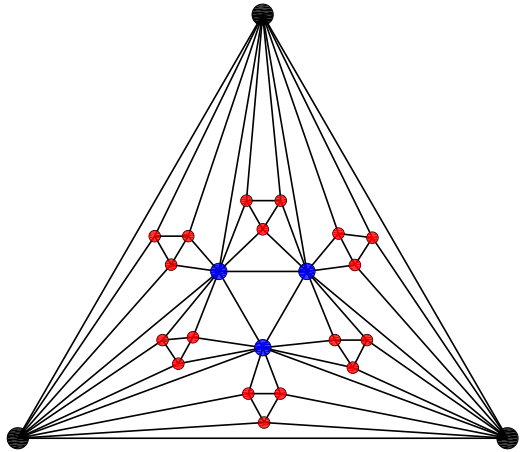}
\end{center}
\caption{\label{apollonian} Illustration of the network
corresponding to the fractal shown in figure~\ref{Sier}.}
\end{figure}

The Sierpinski network is derived from a variation of the Sierpinski
gasket~\cite{Si15}. The Sierpinski gasket variant, shown in
figure~\ref{Sier}, is constructed as follows~\cite{Ha92}. We start
with an equilateral triangle, and we denote this initial
configuration by generation $t=0$. Then in the first generation
$t=1$, we divide the three sides of the equilateral triangle in
three, then join these points and remove the three down pointing
triangles. This forms six copies of the original triangle, and the
procedure is repeated indefinitely for all the new copies. In the
limit of infinite $t$ generations, we get a fractal variant of the
Sierpinski gasket. The Hausdorff dimension of the obtained fractal
is $d_{f}=1+\ln 2/ \ln 3$~\cite{Hu81}. From the fractal, one can
define the Sierpinski network~\cite{ZhZhFaGuZh07}, where vertices
correspond to the removed triangles and two vertices are connected
if the boundaries of the corresponding triangles contact each other.
Note that for uniformity, the three sides of the initial equilateral
triangle at step 0 also correspond to three different vertices.
Figure~\ref{apollonian} shows the network construction process.

\begin{figure}
\begin{center}
\includegraphics[width=0.3\linewidth,trim=100 150 100
80]{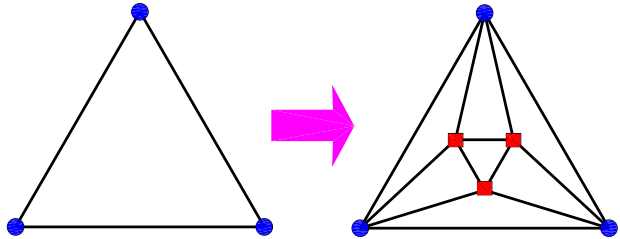}
\end{center}
\caption[kurzform]{\label{iterative} Iterative construction method
for the network. }
\end{figure}

The Sierpinski network can be generated using an iterative
algorithm~\cite{ZhZhFaGuZh07}. We denote the network after $t$
iterations by $F_t$ with $t\geq 0$. Then the network is constructed
as follows: For $t=0$, $F_t$ is a triangle. Next, three nodes are
added into the original triangle. These three new nodes are
connected to each other forming a new triangle, and both ends of
each edge of the new triangle are linked to a node of the original
triangle. Thus $F_1$ is obtained, see figure~\ref{iterative}. For
$t\geq 1$, we can get $F_t$ from $F_{t-1}$. For each of the existing
triangles of $F_{t-1}$ that does not consist of three simultaneously
emerging nodes and has never generated a node before, we define it
an active triangle. We replace each of the active triangles in
$F_{t-1}$ by the connected cluster on the right hand of
figure~\ref{iterative} to obtain $F_t$.

The resulting network presents the typical characteristics of
real-life networks in nature and society~\cite{ZhZhFaGuZh07}. It has
power-law distributions of degree and strength, with exponents
$\gamma_k=2+\frac{\ln2}{\ln3}$ and
$\gamma_s=\frac{3}{2}+\frac{\ln2}{2\ln3}$, respectively. There also
exists a power-law scaling relation between the strength $s$ and
degree $k$ of an individual node, i.e. $s\sim k^2$. On the other
hand, for any individual vertex, its clustering coefficient is
$C(k)= \frac{4}{k}-\frac{1}{k-1}$; so when $k$ is large, $C(k)$ is
approximately inversely proportional to degree $k$. The mean value
$C$ of clustering coefficients of all vertices is very large, which
asymptotically reaches a constant value 0.598. Moreover, the network
is a maximal planar graph.

\section{Rigorous derivation of average distance}

After introducing the Sierpinski network, we now derive analytically
the average distance. 
We represent all the shortest path lengths of network $F_{t}$ as a
matrix in which the entry $d_{ij}$ is the distance between node $i$
and $j$ that is the length of a shortest path joining $i$ and $j$. A
measure of the typical separation between two nodes in $F_{t}$ is
given by the average distance $d_{t}$ defined as the mean of
distances over all pairs of nodes:
\begin{equation}\label{apl01}
d_{t}  = \frac{D_t}{N_t(N_t-1)/2}\,,
\end{equation}
where
\begin{equation}\label{total01}
D_t = \sum_{i \in F_{t},\, j \in F_{t},\, i \neq j} d_{ij}
\end{equation}
denotes the sum of the distances between two nodes over all couples.

\begin{figure}
\begin{center}
\includegraphics[width=8cm]{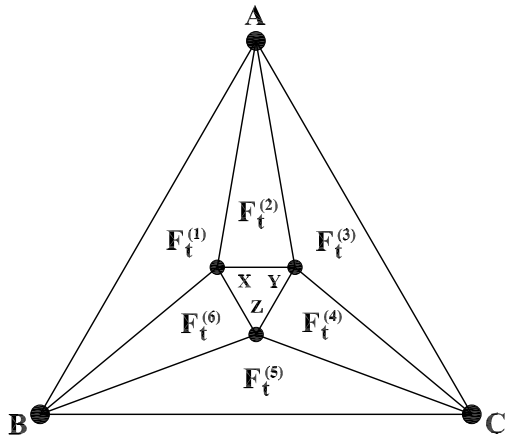}
\caption{Schematic illustration of second construction means of
Sierpinski network. $F_{t+1}$ may be obtained by joining six copies
of $F_t$ denoted as
    $F_t^{(\eta)}$ $(\eta=1,\cdots,6)$, which are
    connected to one another at the edge nodes.} \label{copy}
\end{center}
\end{figure}

\subsection{Recursive equation for total distances}

We continue by exhibiting the procedure of determining the total
distance and present the recurrence formula, which allows us to
obtain $D_{t+1}$ of the $t+1$ generation from $D_{t}$ of the $t$
generation. The Sierpinski network $F_{t}$ has a self-similar
structure that allows one to calculate $D_{t}$
analytically~\cite{HiBe06}. As shown in figure~\ref{copy}, network
$F_{t+1}$ may be obtained by joining at six edge nodes (i.e., $A$,
$B$, $C$, $X$, $Y$, and $Z$) six copies of $F_{t}$ that are labeled
as $F_{t}^{(1)}$, $\cdots$, $F_{t}^{(6)}$~\cite{Bobe05}. From this
we can obtain the recursion relation
\begin{equation}\label{order01}
  N_{t+1} =6\,N_{t}-12,
\end{equation}
for the network order $N_{t}$, which is the number of nodes in the
graph of generation $t$. This recursion, coupled with $N_{0}=3$,
yields
\begin{equation}\label{order02}
  N_{t}=\frac{3\times 6^{t}+12}{5}
\end{equation}
as previously obtained in~\cite{ZhZhFaGuZh07}.

According to the second construction method, the total distance
$D_{t+1}$ satisfies the recursion relation
\begin{equation}\label{total02}
  D_{t+1} = 6\, D_t + \Delta_t-6,
\end{equation}
where $\Delta_t$ is the sum over all shortest path length whose
endpoints are not in the same $F_t^{(\eta)}$ branch. The last term
-6 on the right-hand side of Eq.~(\ref{total02}) compensates for the
overcounting of certain paths: the shortest path between $A$ and
$X$, with length 1, is included in both $F_{t}^{(1)}$ and
$F_{t}^{(2)}$. Similarly the shortest path between $A$ and $Y$, the
shortest path between $B$ and $X$, the shortest path between $B$ and
$Z$, the shortest path between $C$ and $Y$, and the shortest path
between $C$ and $Z$, are all computed twice. To determine $D_t$, all
that is left is to calculate $\Delta_t$.

\subsection{Definition of crossing distance}

In order to compute $\Delta_t$, we classify the nodes in $F_{t+1}$
into two categories: the six edge nodes (such as $A$, $B$, $C$, $X$,
$Y$, and $Z$ in figure~\ref{copy}) are called hub nodes, while the
other nodes are named non-hub nodes. Thus $\Delta_t$, named the
crossing distance, can be obtained by summing the following path
length that are not included in the distance of node pairs in
$F_{t}^{(\eta)}$: length of the shortest paths between non-hub and
non-hub nodes, length of the shortest paths between hub and non-hub
nodes, and length of the shortest paths between hub nodes (i.e.,
$d_{AZ}$, $d_{BY}$, and $d_{CZ}$).

Denote $\Delta_t^{\alpha,\beta}$ as the sum of all shortest paths
between non-hub nodes, whose endpoints are in $F_t^{(\alpha)}$ and
$F_t^{(\beta)}$, respectively. That is to say,
$\Delta_t^{\alpha,\beta}$ rules out the paths with endpoint at the
hub nodes belonging to $F_t^{(\alpha)}$ or $F_t^{(\beta)}$. For
example, each path contributed to $\Delta_t^{1,2}$ does not end at
node $A$, $B$, $X$ or $Y$. On the other hand, let $\Omega_t^{\eta}$
be the set of non-hub nodes in $F_{t}^{(\eta)}$. Then the total sum
$\Delta_t$ is given by
\begin{eqnarray}\label{cross01}
\Delta_t =\Delta_t^{1,2} &+& \Delta_t^{1,3} + \Delta_t^{1,4}+
\Delta_t^{1,5}+ \Delta_t^{1,6}+ \Delta_t^{2,3}+ \Delta_t^{2,4}+
\Delta_t^{2,5}+ \Delta_t^{2,6}+ \Delta_t^{3,4}+ \Delta_t^{3,5}+
\Delta_t^{3,6}+\Delta_t^{4,5} \nonumber\\
&+&\Delta_t^{4,6}+\Delta_t^{5,6}+\sum_{j \in
\Omega_t^{4}}d_{Aj}+\sum_{j \in \Omega_t^{5}}d_{Aj}+\sum_{j \in
\Omega_t^{6}}d_{Aj}+\sum_{j \in \Omega_t^{2}}d_{Bj}+\sum_{j \in
\Omega_t^{3}}d_{Bj}+\sum_{j \in \Omega_t^{4}}d_{Bj}\nonumber\\
&+&\sum_{j \in \Omega_t^{1}}d_{Cj}+\sum_{j \in
\Omega_t^{2}}d_{Cj}+\sum_{j \in \Omega_t^{6}}d_{Cj}+\sum_{j \in
\Omega_t^{3}}d_{Xj}+\sum_{j \in \Omega_t^{4}}d_{Xj}+\sum_{j \in
\Omega_t^{5}}d_{Xj}+\sum_{j \in \Omega_t^{1}}d_{Yj}+\sum_{j \in
\Omega_t^{5}}d_{Yj}\nonumber\\
&+&\sum_{j \in \Omega_t^{6}}d_{Yj}+\sum_{j \in
\Omega_t^{1}}d_{Zj}+\sum_{j \in \Omega_t^{2}}d_{Zj}+\sum_{j \in
\Omega_t^{3}}d_{Zj}+d_{AZ}+d_{BY}+d_{CX}
\end{eqnarray}

By symmetry, $\Delta_t^{1,2} =\Delta_t^{1,6} = \Delta_t^{2,3}+
\Delta_t^{3,4}=\Delta_t^{4,5}=\Delta_t^{5,6}$, $\Delta_t^{1,3}
=\Delta_t^{1,5}=\Delta_t^{2,4}=\Delta_t^{2,6}=\Delta_t^{3,5}=\Delta_t^{4,6}$,
$\Delta_t^{1,4}=\Delta_t^{2,5}=\Delta_t^{3,6}$, $\sum_{j \in
\Omega_t^{4}}d_{Aj}=\sum_{j \in \Omega_t^{5}}d_{Aj}=\sum_{j \in
\Omega_t^{6}}d_{Aj}=\sum_{j \in \Omega_t^{2}}d_{Bj}=\sum_{j \in
\Omega_t^{3}}d_{Bj}=\sum_{j \in \Omega_t^{4}}d_{Bj}=\sum_{j \in
\Omega_t^{1}}d_{Cj}=\sum_{j \in \Omega_t^{2}}d_{Cj}=\sum_{j \in
\Omega_t^{6}}d_{Cj}=\sum_{j \in \Omega_t^{3}}d_{Xj}=\sum_{j \in
\Omega_t^{4}}d_{Xj}=\sum_{j \in \Omega_t^{5}}d_{Xj}=\sum_{j \in
\Omega_t^{1}}d_{Yj}=\sum_{j \in \Omega_t^{5}}d_{Yj}=\sum_{j \in
\Omega_t^{6}}d_{Yj}=\sum_{j \in \Omega_t^{1}}d_{Zj}=\sum_{j \in
\Omega_t^{2}}d_{Zj}=\sum_{j \in \Omega_t^{3}}d_{Zj}$, and
$d_{AZ}=d_{BY}=d_{CX}=2$, so Eq.~(\ref{cross01}) can be simplified
as
\begin{equation}\label{cross02}
\Delta_t
=6\,\Delta_t^{1,2}+6\,\Delta_t^{1,3}+3\,\Delta_t^{1,4}+18\,\sum_{j
\in \Omega_t^{4}}d_{Aj}+6.
\end{equation}
Having $\Delta_t$ in terms of the quantities of $\Delta_t^{1,2}$,
$\Delta_t^{1,3}$, $\Delta_t^{1,4}$, and $\sum_{j \in
\Omega_t^{4}}d_{Aj}$, the next step is to explicitly determine these
quantities.

\begin{figure}
\begin{center}
\includegraphics[width=14cm]{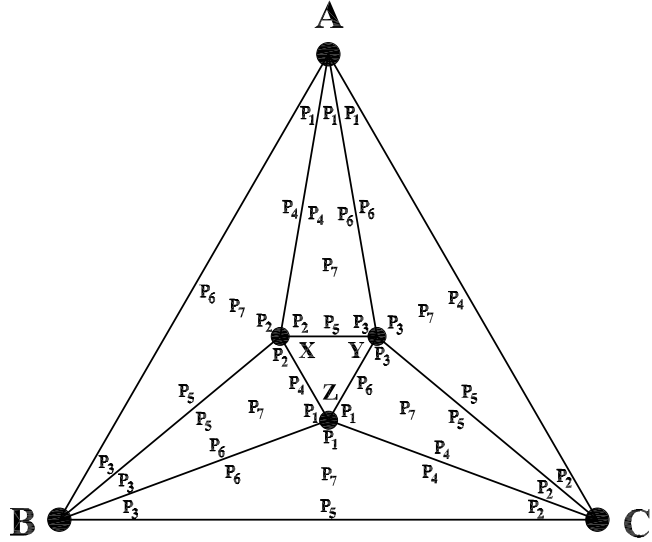}
\caption{Illustration of the classification of interior nodes in
$F_t^{(\eta)}$ $(\eta=1,\cdots,6)$, from which we can derive
recursively the classification of interior nodes in network
$F_{t+1}$.}\label{class}
\end{center}
\end{figure}

\subsection{Classification of interior nodes}

To calculate the crossing distance $\Delta_t^{1,2}$,
$\Delta_t^{1,3}$, $\Delta_t^{1,4}$, and $\sum_{j \in
\Omega_t^{4}}d_{Aj}$, we classify interior nodes in network
$F_{t+1}$ into seven different parts according to their shortest
path lengths to each of the three hub nodes (i.e. $A$, $B$, $C$) of
the peripheral triangle $\triangle ABC$. Notice that nodes $A$, $B$,
$C$ themselves are not partitioned into any of the seven parts
represented as $P_{1}$, $P_{2}$, $P_{3}$, $P_{4}$, $P_{5}$, $P_{6}$,
and $P_{7}$, respectively. The classification of nodes is shown in
figure~\ref{class}. For any interior node $v$, we denote the
shortest path lengths from $v$ to $A$, $B$, $C$ as $a$, $b$, and
$c$, respectively. By construction, $a$, $b$, $c$ can differ by at
most $1$ since vertices $A$, $B$, $C$ are adjacent. Then the
classification function $class(v)$ of node $v$ is defined to be

\begin{equation}\label{classification}
class(v)=\left\{
\begin{array}{lc}
{\displaystyle{P_{1}}}
& \quad \hbox{for}\ a<b=c,\\
{\displaystyle{P_{2}}}
& \quad \hbox{for}\ b<a=c,\\
{\displaystyle{P_{3}}}
& \quad \hbox{for}\ c<a=b,\\
{\displaystyle{P_{4}}}
& \quad \hbox{for}\ a=c<b,\\
{\displaystyle{P_{5}}}
& \quad \hbox{for}\ a=b<c,\\
{\displaystyle{P_{6}}}
& \quad \hbox{for}\ b=c<a,\\
{\displaystyle{P_{7}}}
& \quad \hbox{for}\ a=b=c.\\
\end{array} \right.
\end{equation}

It should be mentioned that the definition of node classification is
recursive. For instance, class $P_{1}$ and $P_{4}$ in $F_t^{(1)}$
belong to class $P_{1}$ in $F_{t+1}$, class $P_{3}$ and $P_{5}$ in
$F_t^{(1)}$ belong to class $P_{2}$ in $F_{t+1}$, class $P_{2}$,
$P_{6}$, and $P_{7}$ in $F_t^{(1)}$ belong to class $P_{5}$ in
$F_{t+1}$. Since the three nodes $A$, $B$, and $C$ are symmetrical,
in the Sierpinski network we have the following equivalent relations
from the viewpoint of class cardinality: classes $P_{1}$, $P_{2}$,
and $P_{3}$ are equivalent to one another, and it is the same with
classes $P_{4}$, $P_{5}$, and $P_{6}$. We denote the number of nodes
in network $F_{t}$ that belong to class $P_{1}$ as $N_{t,P_{1}}$,
the number of nodes in class $P_{2}$ as $N_{t,P_{2}}$, and so on. By
symmetry, we have $N_{t,P_{1}}=N_{t,P_{2}}=N_{t,P_{3}}$ and
$N_{t,P_{4}}=N_{t,P_{5}}=N_{t,P_{6}}$. Therefore in the following
computation we will only consider $N_{t,P_{1}}$, $N_{t,P_{4}}$, and
$N_{t,P_{7}}$. It is easy to conclude that
\begin{align}
N_{t} &= N_{t,P_{1}}+N_{t,P_{2}}+N_{t,P_{3}}+N_{t,P_{4}}+N_{t,P_{5}}+N_{t,P_{6}}+N_{t,P_{7}}+3 \nonumber\\
 &=3\,N_{t,P_{1}} + 3\,N_{t,P_{4}} + N_{t,P_{7}}+3.
\end{align}
Considering the self-similar structure of Sierpinski network, we can
easily know that at time $t+1$, the quantities $N_{t+1,P_{1}}$,
$N_{t+1,P_{4}}$, and $N_{t+1,P_{7}}$ evolve according to the
following recursive equations
\begin{eqnarray}\label{Np01}
\left\{
\begin{array}{ccc}
N_{t+1,P_{1}} &=& 3\,N_{t,P_{1}} + 4\,N_{t,P_{4}}+N_{t,P_{7}}\,, \\
N_{t+1,P_{4}} &=& 3\,N_{t,P_{1}} + N_{t,P_{4}}+N_{t,P_{7}}+1\,, \\
N_{t+1,P_{7}} &=& 3\,N_{t,P_{4}}\,, \\
 \end{array}
 \right.
\end{eqnarray}
where we have used the equivalent relations
$N_{t,P_{1}}=N_{t,P_{2}}=N_{t,P_{3}}$ and
$N_{t,P_{4}}=N_{t,P_{5}}=N_{t,P_{6}}$.  With the initial condition
$N_{2,P_{1}}=4$, $N_{2,P_{4}}=2$, and $N_{2,P_{7}}=3$, we can solve
the recursive equation~(\ref{Np01}) to obtain

\begin{eqnarray}\label{Np02}
\left\{
\begin{array}{ccc}
N_{t,P_{1}} &=& \frac{1}{240} \left[-112+25 \times(-2)^t+27 \times 6^t \right]\,, \\
N_{t,P_{4}} &=& \frac{1}{120} \left[16-25 \times(-2)^t+9\times 6^t \right]\,, \\
N_{t,P_{7}} &=& =\frac{1}{80} \left[32+25 \times(-2)^t+3\times
6^t\right]
\,. \\
\end{array}
\right.
\end{eqnarray}

For a node $v$ in network $F_{t+1}$, we are also interested in the
smallest value of the shortest path length from $v$ to any of the
three peripheral hub nodes $A$, $B$, and $C$. We denote the shortest
distance as $f_v$, which can be defined to be
\begin{equation}\label{fv01}
f_v = min(a,b,c).
\end{equation}

Let $d_{t,P_{1}}$ denote the sum of $f_v$ of all nodes belonging to
class $P_{1}$ in network $F_{t}$. Analogously, we can also define
the quantities $d_{t,P_{2}}$, $d_{t,P_{3}}$, $\cdots$,
$d_{t,P_{7}}$. Again by symmetry, we have
$d_{t,P_{1}}=d_{t,P_{2}}=d_{t,P_{3}}$,
$d_{t,P_{4}}=d_{t,P_{5}}=d_{t,P_{6}}$, and $d_{t,P_{1}}$,
$d_{t,P_{4}}$, $d_{t,P_{7}}$ can be written recursively as follows:

\begin{eqnarray}\label{dp01}
\left\{
\begin{array}{ccc}
d_{t+1,P_{1}} &=& 3\,d_{t,P_{1}} +4\, d_{t,P_{4}}+ d_{t,P_{7}}\,, \\
d_{t+1,P_{4}} &=& 3\,d_{t,P_{1}}+d_{t,P_{4}} + d_{t,P_{7}}+3\,N_{t,P_{1}}+1\,, \\
d_{t+1,P_{7}} &=& 3\,(d_{t,P_{4}} + N_{t,P_{4}})\,. \\
 \end{array}
 \right.
\end{eqnarray}

Substituting Eq.~(\ref{Np02}) into Eq.~(\ref{dp01}), and considering
the initial condition $d_{2,P_{1}}=4$, $d_{2,P_{4}}=2$, and
$d_{2,P_{7}}=6$, Eq.~(\ref{dp01}) is solved inductively
\begin{eqnarray}\label{dp02}
\left\{
\begin{array}{ccc}
d_{t,P_{1}}&=& \frac{1}{25600}[2048+327\times 6^t+990t \times 6^{t}+25(33+70 t)2^t e^{i \pi  t} ]\,,\\
d_{t,P_{4}}&=& \frac{1}{12800}\left[-2075 (-2)^t-1024 (-1)^{2 t}+699\times 6^{t}+10t \left(-175\times (-2)^t+33\times 6^{t}\right) \right]\,,\\
d_{t,P_{7}}&=& \frac{1}{25600}[4096+1329\times 6^t +330t\times 6^{t}
+25 (359+210 t) 2^t\, e^{i \pi  t}]\,.
 \end{array}
 \right.
\end{eqnarray}

\subsection{Calculation of crossing distances}

Having obtained the quantities $N_{t,P_{i}}$ and $d_{t,P_{i}}$
($i=1,2,\cdots, 7$), we now begin to determine the crossing distance
$\Delta_t^{1,2}$, $\Delta_t^{1,3}$, $\Delta_t^{1,4}$, and $\sum_{j
\in \Omega_t^{4}}d_{Aj}$ expressed as a function of $N_{t,P_{i}}$
and $d_{t,P_{i}}$. Here we only give the computation details of
$\Delta_t^{1,2}$, while the computing processes of $\Delta_t^{1,3}$,
$\Delta_t^{1,4}$, and $\sum_{j \in \Omega_t^{4}}d_{Aj}$ are similar.
For convenience of computation, we use $\Gamma_t^{\eta,i}$ to denote
the set of interior nodes belonging to class $P_i$ in
$F_{t}^{(\eta)}$. Then $\Delta_t^{1,2}$ can be written as

\begin{eqnarray}\label{cross04}
  \Delta_t^{1,2} = \sum_{\substack{u \in \Gamma_t^{1,1},\,\,v\in
      F_t^{(2)}\\ v \ne A, X, Y }} d_{uv}&+&\sum_{\substack{u \in \Gamma_t^{1,2},\,\,v\in
      F_t^{(2)}\\ v \ne A, X, Y }} d_{uv}+\sum_{\substack{u \in \Gamma_t^{1,3},\,\,v\in
      F_t^{(2)}\\ v \ne A, X, Y }} d_{uv}+\sum_{\substack{u \in \Gamma_t^{1,4},\,\,v\in
      F_t^{(2)}\\ v \ne A, X, Y }} d_{uv}\nonumber \\
      &+&\sum_{\substack{u \in \Gamma_t^{1,5},\,\,v\in
      F_t^{(2)}\\ v \ne A, X, Y }} d_{uv}+\sum_{\substack{u \in \Gamma_t^{1,6},\,\,v\in
      F_t^{(2)}\\ v \ne A, X, Y }} d_{uv}+\sum_{\substack{u \in \Gamma_t^{1,7},\,\,v\in
      F_t^{(2)}\\ v \ne A, X, Y }} d_{uv}.
\end{eqnarray}
The seven terms on the right-hand side of Eq.~(\ref{cross04}) are
represented consecutively as $\delta_t^i$ ($i=1,2,\cdots, 7$). Next
we will calculate the quantities $\delta_t^i$. By symmetry,
$\delta_t^1=\delta_t^2$, $\delta_t^5=\delta_t^6$. Therefore, we need
only to compute $\delta_t^1$, $\delta_t^3$, $\delta_t^4$,
$\delta_t^5$ and $\delta_t^7$. Firstly, we evaluate $\delta_t^1$. By
definition,
\begin{eqnarray}\label{cross05}
  \delta_t^1 &=&\sum_{\substack{u \in \Gamma_t^{1,1},\,\,v\in
      F_t^{(2)}\\ v \ne A, X, Y }} d_{uv}\nonumber \\
      &=& \sum_{u \in \Gamma_t^{1,1},\, v\in
      \Gamma_t^{2,1}\bigcup
      \Gamma_t^{2,4}\bigcup\Gamma_t^{2,6}\bigcup\Gamma_t^{2,7}}
      (d_{uA}+d_{Av})+\sum_{u \in \Gamma_t^{1,1},\, v\in
      \Gamma_t^{2,3}}(d_{uA}+d_{AY}+d_{Yv})\nonumber \\&\quad&+\sum_{u \in \Gamma_t^{1,1},\, v\in
      \Gamma_t^{2,2}\bigcup
      \Gamma_t^{2,5}}(d_{uA}+d_{AX}+d_{Xv})\nonumber \\
      &=&N_{t,P_{1}}(3d_{t,P_{1}}+3d_{t,P_{4}}+d_{t,P_{7}}+2N_{t,P_{1}}+N_{t,P_{4}})+d_{t,P_{1}}(3N_{t,P_{1}}+3N_{t,P_{4}}+N_{t,P_{7}}).
\end{eqnarray}
Proceeding similarly, we obtain
\begin{eqnarray}\label{cross06}
  \delta_t^3
  =N_{t,P_{1}}(3d_{t,P_{1}}+3d_{t,P_{4}}+d_{t,P_{7}}+4N_{t,P_{1}}+3N_{t,P_{4}}+N_{t,P_{7}})+d_{t,P_{1}}(3N_{t,P_{1}}+3N_{t,P_{4}}+N_{t,P_{7}}),
\end{eqnarray}
\begin{eqnarray}\label{cross07}
  \delta_t^4
  =N_{t,P_{2}}(3d_{t,P_{1}}+3d_{t,P_{4}}+d_{t,P_{7}}+N_{t,P_{1}})+d_{t,P_{2}}(3N_{t,P_{1}}+3N_{t,P_{4}}+N_{t,P_{7}}),
\end{eqnarray}
\begin{eqnarray}\label{cross08}
  \delta_t^5
  =N_{t,P_{2}}(3d_{t,P_{1}}+3d_{t,P_{4}}+d_{t,P_{7}}+2N_{t,P_{1}}+N_{t,P_{4}})+d_{t,P_{2}}(3N_{t,P_{1}}+3N_{t,P_{4}}+N_{t,P_{7}}),
\end{eqnarray}
and
\begin{eqnarray}\label{cross09}
  \delta_t^7
  =N_{t,P_{3}}(3d_{t,P_{1}}+3d_{t,P_{4}}+d_{t,P_{7}}+N_{t,P_{1}})+d_{t,P_{7}}(3N_{t,P_{1}}+3N_{t,P_{4}}+N_{t,P_{7}}).
\end{eqnarray}
With the obtained results for $\delta_t^i$, we have
\begin{eqnarray}\label{cross10}
  \Delta_t^{1,2}
  &=&2(3d_{t,P_{1}}+3d_{t,P_{4}}+d_{t,P_{7}})(3N_{t,P_{1}}+3N_{t,P_{4}}+N_{t,P_{7}})+N_{t,P_{1}}(3N_{t,P_{1}}+3N_{t,P_{4}}+N_{t,P_{7}})\nonumber \\
&\quad&+2(N_{t,P_{1}}+N_{t,P_{4}})(2N_{t,P_{1}}+N_{t,P_{4}})+N_{t,P_{1}}(N_{t,P_{4}}+N_{t,P_{7}})+(N_{t,P_{1}})^2.
\end{eqnarray}
Analogously, we find
\begin{eqnarray}\label{cross11}
  \Delta_t^{1,3}
  &=&2(3d_{t,P_{1}}+3d_{t,P_{4}}+d_{t,P_{7}})(3N_{t,P_{1}}+3N_{t,P_{4}}+N_{t,P_{7}})+2(N_{t,P_{1}})^2+2N_{t,P_{1}}(3N_{t,P_{1}}+3N_{t,P_{4}}+N_{t,P_{7}})\nonumber \\
&\quad&+N_{t,P_{4}}(3N_{t,P_{1}}+3N_{t,P_{4}}+N_{t,P_{7}})+(N_{t,P_{1}}+2N_{t,P_{4}}+N_{t,P_{7}})(2N_{t,P_{1}}+N_{t,P_{4}}),
\end{eqnarray}
\begin{eqnarray}\label{cross12}
  \Delta_t^{1,4}
  =2(3d_{t,P_{1}}+3d_{t,P_{4}}+d_{t,P_{7}})(3N_{t,P_{1}}+3N_{t,P_{4}}+N_{t,P_{7}})+(3N_{t,P_{1}}+3N_{t,P_{4}}+N_{t,P_{7}})^2+3(N_{t,P_{1}})^2,
\end{eqnarray}
and
\begin{equation}\label{cross13}
 \sum_{j \in \Omega_t^{4}}d_{Aj}
  =(3d_{t,P_{1}}+3d_{t,P_{4}}+d_{t,P_{7}})+(3N_{t,P_{1}}+3N_{t,P_{4}}+N_{t,P_{7}})+N_{t,P_{1}}.
\end{equation}

Substituting Eqs.~(\ref{cross10}), (\ref{cross11}), (\ref{cross12}),
and (\ref{cross13}) into Eq. (\ref{cross02}), we the final
expression for cross distances $\Delta_t$,
\begin{eqnarray}\label{cross14}
\Delta_t =\frac{1}{19200}\big[&-&17920+95040\times 6^{t}
+160380\times 6^{2 t}+50\times 2^{t} (-32+27\times6^t) e^{i \pi t}+
44550t\times6^{2t} \nonumber \\
&+&e^{2 i \pi  t} (6875\times 4^{t}-82944\times 6^{t} +8019\times
6^{2t}+26730t\times 6^{2t})\big].
\end{eqnarray}

\subsection{Rigorous result of average distance}

With the above-obtained results and recursion relations, we now
readily calculate the sum of the shortest path lengths between all
pairs of nodes. Inserting Eq.~(\ref{cross14}) into
Eq.~(\ref{total02}) and using the initial condition $D_{2} =555$,
Eq.~(\ref{total02}) is solved inductively,

\begin{eqnarray}\label{total04}
D_{t} =\frac{1}{192000}\Big[266240-34375\times 4^{t}&+&2000\times
(-2)^t +225264\times 6^{t} -750 (-12)^t\nonumber\\
&+&27621\times 6^{2t} + 20160t\times 6^{t}  +23760t\times 6^{2
t}\Big].
\end{eqnarray}

\begin{figure}
\begin{center}
\includegraphics[width=.20\linewidth,trim=120 20 120 30]{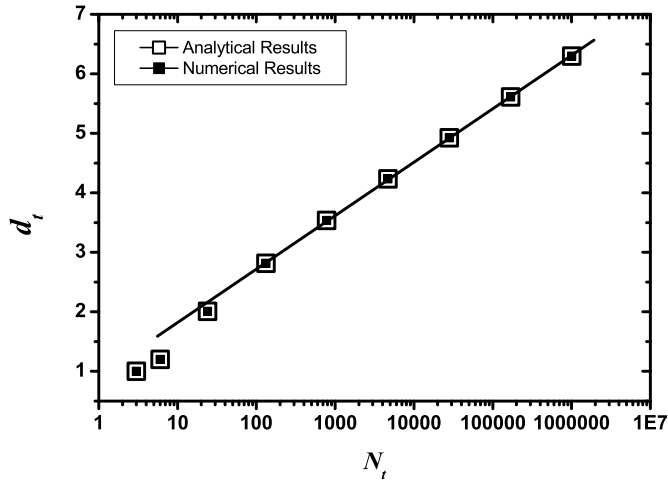}
\end{center}
\caption[kurzform]{\label{AveDis} Average distance $d_{t}$ versus
network order $N_{t}$ on a semi-logarithmic scale. The solid line
serves as guide to the eye.}
\end{figure}

Substituting Eq.~(\ref{total04}) into  Eq.~(\ref{apl01}) yields the
exactly  analytic expression for average distance
\begin{eqnarray}\label{apl02}
d_{t} =\frac{1}{34560
(84+57\times6^t+9\times6^{2t})}&\quad&\Big[266240-34375\times
4^{t}+2000\times (-2)^t +225264\times 6^{t}\nonumber\\ &-&750
(-12)^t +27621\times 6^{2t} + 20160t\times 6^{t}  +23760t\times 6^{2
t}\Big],
\end{eqnarray}
In the large $t$ limit, $d_{t}\sim t$, while the network order $N_t
\sim 6^t$ which is obvious from Eq.~(\ref{order02}). Thus, the
average distance grows logarithmically with increasing order of the
network. This scaling is consistent with the speculation
in~\cite{ZhZhFaGuZh07} based on computer simulations. We have also
checked our analytic result provide by Eq.~(\ref{apl02}) against
numerical calculations for different network order up to $t=8$ which
corresponds to $N_{8}=1007772$. In all the cases we obtain a
complete agreement between our theoretical formula and the results
of numerical investigation, see figure~\ref{AveDis}.

Recently, it has been suggested that for random scale-free networks
with degree exponent $\gamma_k <3$ and network order $N$, their
average distance $d(N)$ behaves as a double logarithmic scaling with
$N$: $d(N)\sim \ln\ln N$~\cite{CoHa03,ChLu02}. However, for
deterministic Sierpinsiki network, in despite of the fact that its
degree exponent $\gamma_k=2+\frac{\ln2}{\ln3} <3$, its average
distance scales as a logarithmic scaling with network order, showing
a obvious difference from that of the stochastic scale-free
counterparts.


\section{Conclusion}

Average distance plays an important role in the characterization of
the internal structure of a network, and has a profound impact on a
variety of dynamical processes on the network. In this article, we
have obtained rigorously the solution for the average distance of a
deterministic Sierpinski network. We have explicitly shown that in
the limit of infinite network order, the average distance of
Sierpinski network scales logarithmically with the number of network
nodes, verifying our previously suggested scaling obtained through
simulations~\cite{ZhZhFaGuZh07}. Our findings display that the
scaling of the average distance for deterministic Sierpinski network
is strikingly distinct from the counterpart of stochastic scale-free
networks~\cite{CoHa03,ChLu02}. This disparity of the scaling for
average distance between the deterministic Sierpinsk network and
random scale-free networks is worth studying in future.

\section*{Acknowledgment}

This research was supported by the National Basic Research Program
of China under grant No. 2007CB310806, the National Natural Science
Foundation of China under Grant Nos. 60704044, 60873040 and
60873070, Shanghai Leading Academic Discipline Project No. B114, and
the Program for New Century Excellent Talents in University of China
(NCET-06-0376).


\begin{references}

\bibitem{CoRoTrVi07}
L. da. F. Costa, F. A. Rodrigues, G. Travieso, and P. R. V. Boas,
Adv. Phys. {\bf 56}, 167 (2007).

\bibitem{BaAl99} A.-L. Barab\'asi and R. Albert,
       Science {\bf 286}, 509 (1999).

\bibitem{WaSt98} D.J. Watts and H. Strogatz,
        Nature (London) {\bf 393}, 440 (1998).

\bibitem{Newman02}
M. E. J. Newman,
Phys. Rev. Lett. {\bf 89}, 208701 (2002).

\bibitem{GiNe02}
M. Girvan and M. E. J. Newman, Proc. Natl. Acad. Sci. U.S.A. {\bf
99}, 7821 (2002).

\bibitem{MiShItKaChAl02}
R. Milo, S. Shen-Orr, S. Itzkovitz, N. Kashtan, D. Chklovskii, and
U. Alon, Science {\bf 298}, 824 (2002).

\bibitem{SoHaMa05}
 C. Song, S. Havlin, H. A. Makse,
Nature {\bf 433}, 392 (2005).

\bibitem{XiXiWaWa08}
Y. Xiao, M. Xiong, W. Wang, and H. Wang, Phys. Rev. E {\bf 77},
066108 (2008).

\bibitem{ChLu02}
 F. Chung and L. Lu, Proc. Natl. Acad. Sci. U.S.A. {\bf 99}, 15879 (2002).

\bibitem{CoHa03}
R. Cohen and S. Havlin, Phys. Rev. Lett. {\bf 90}, 058701 (2003).

\bibitem{SoHaMa06}
 C. Song, S. Havlin, H. A. Makse,
Nature Phys. {\bf 2}, 275 (2006).

\bibitem{ZhZhZo07}
Z. Z. Zhang, S. G. Zhou, and T. Zou, Eur. Phys. J. B {\bf 56}, 259
(2007).

\bibitem{XiMaiWaXiWa08}
Y. Xiao, B. D. MacArthur, H. Wang, M. Xiong, and W. Wang, Phys. Rev.
E {\bf 78}, 046102 (2008).

\bibitem{YaZhHuFuWa06}
G. Yan, T. Zhou, B. Hu, Z. Q. Fu, and B. H. Wang, Phys. Rev. E {\bf
73}, 046108 (2006).

\bibitem{ZhCoFeRaRoZh08}
Z. Z. Zhang, F. Comellas, G. Fertin, A. Raspaud, L. L. Rong, and S. G. Zhou,
J. Phys. A: Math. Theor. {\bf 41}, 035004 (2008).

\bibitem{ZhZhZoCh08}
Z. Z. Zhang, S. G. Zhou, T. Zou, and G. S. Chen, J. Stat. Mech.:
Theory Exp. {\bf P09008} (2008).


\bibitem{DoMeSa03}
S. N. Dorogovtsev, J. F. F. Mendes, and A.N. Samukhin, Nucl. Phys.
{\bf 653}, 307 (2003).

\bibitem{FrFrHo04}
A. Fronczak, P. Fronczak, and J. A. Ho{\l}yst, Phys. Rev. E {\bf
70}, 056110 (2004).

\bibitem{Lolo03}
W. S. Lovejoy, C. H. Loch, Soc. Netw. {\bf 25}, 333 (2003).

\bibitem{HoSiFrFrSu05}
J. A. Ho{\l}yst, J. Sienkiewicz, A. Fronczak, P. Fronczak, and K.
Suchecki, Phys. Rev. E {\bf 72}, 026108 (2005).


\bibitem{DoMeOl06}
S. N. Dorogovtsev, J. F. F. Mendes, and J. G. Oliveira, Phys. Rev. E
{\bf 73}, 056122 (2006).

\bibitem{ZhZhChYiGu08}
Z. Z. Zhang, S. G. Zhou, L. C. Chen, M. Yin, and J. H. Guan, J.
Phys. A: Math. Theor. {\bf 41}, 485102 (2008).


\bibitem{AlBa02} R. Albert and A.-L. Barab\'asi,
       Rev. Mod. Phys. {\bf 74}, 47 (2002).

\bibitem{DoMe02} S. N. Dorogovtsev and J.F.F. Mendes,
Adv. Phys. {\bf 51}, 1079 (2002).

\bibitem{Ne03} M. E. J. Newman,
SIAM Review {\bf 45}, 167 (2003).

\bibitem{BoLaMoChHw06}
S. Boccaletti, V. Latora, Y. Moreno, M. Chavez, and D.-U. Hwanga,
Phy. Rep. {\bf 424}, 175 (2006).


\bibitem{ZhZhFaGuZh07}
Z.Z. Zhang, S. G. Zhou, L. J. Fang, J. H. Guan, and Y. C. Zhang, EPL
{\bf 79}, 38007 (2007).


\bibitem{BaRaVi01} A.-L. Barab\'asi, E. Ravasz, and T. Vicsek,
          Physica A  {\bf 299}, 559 (2001).

\bibitem{DoGoMe02} S.N. Dorogovtsev, A.V. Goltsev, and J.F.F. Mendes,
          Phys. Rev. E {\bf 65}, 066122 (2002).

\bibitem{JuKiKa02} S. Jung, S. Kim, and B. Kahng,
        Phys. Rev. E {\bf 65}, 056101 (2002).

\bibitem{RaSoMoOlBa02}
E. Ravasz, A.L. Somera, D. A. Mongru, Z. N. Oltvai, and A.-L.
Barab\'asi, Science {\bf 297}, 1551 (2002).

\bibitem{AnHeAnSi05}
J.S. Andrade Jr., H.J. Herrmann, R.F.S. Andrade and L.R.da Silva,
Phys. Rev. Lett. {\bf 94}, 018702 (2005).


\bibitem{Hi07}
M. Hinczewsk, Phys. Rev. E {\bf 75}, 061104 (2007).


\bibitem{BoGoGu08}
S. Boettcher, B. Gon\c calves, and H. Guclu, J. Phys. A: Math.
Theor. {\bf 41}, 252001 (2008).

\bibitem{Si15}
 W. Sierpinski, Comptes Rendus (Paris) {\bf 160}, 302
(1915).

\bibitem{Ha92}
B.M. Hambly,
Probab. Theory Related Fields {\bf 94}, 1 (1992).

\bibitem{Hu81}
S. Hutchinson,
Indiana Univ. Math. J. {\bf 30}, 713 (1981).

\bibitem{HiBe06}
M. Hinczewski and A. N. Berker, Phys. Rev. E {\bf 73}, 066126
(2006).

\bibitem{Bobe05}
E. M. Bollt, D. ben-Avraham, New J. Phys. {\bf 7}, 26 (2005).


\end{references}
\end{document}